# Electronic and Magnetic Properties of Sr and Ca Doped Lanthanum Manganites from First-Principles

M. –H. Tsai *, Y. –H. Tang, H. Chou, and W. T. Wu

*Department of Physics, National Sun Yat-Sen University, Kaohsiung 80424 Taiwan*

The complicated electronic, magnetic, and colossal magnetoresistant (CMR) properties of Sr and Ca doped lanthanum manganites can be understood by spin-polarized first-principles calculations. The electronic properties can be attributed to a detailed balancing between Sr and Ca induced metal-like O $2p$ and majority-spin ($\uparrow$-spin) Mn $e_g$ delocalized states and the insulator-like minority-spin ($\downarrow$-spin) Mn $t_{2g}$ band near the Fermi level ($E_F$). The magnetic properties can be attributed to a detailed balancing between O mediated antiferromagnetic superexchange and delocalized $\uparrow$-spin Mn $e_g$-state mediated ferromagnetic spin-spin couplings. While CMR can be attributed to the lining up of magnetic domains trigged by the applied magnetic field, which suppresses the trapping ability of the empty Mn $t_{2g}$ states that resists the motion of conducting Mn $\uparrow$-spin $e_g$ electrons.

## I. INTRODUCTION

Rare-earth manganites doped with alkaline-earth metals are important materials of fundamental and technological interests[1-12]. These materials exhibit colossal magnetoresistance (CMR), which is potentially applicable for magnetic devices. They also exhibit interesting properties such as antiferromagnetism-ferromagnetism and insulator-like-metal-like transitions. Based on the decrease of resistivity, $\rho$, with temperature, T, and the observance of an optical energy gap[13], the antiferromagnetic low-temperature $LaMnO_3$ has been described as an insulator. Highly Sr or Ca doped $LaMnO_3$, i.e. $La_{1-x}Sr_xMnO_3$ and $La_{1-x}Ca_xMnO_3$ with a large x, are antiferromagnetic and have a decreasing $\rho$ with T, so is also described as an insulator, despite that the decrease of $\rho$ is not exponential and $\rho$ has an order of only $10^2$ $\Omega$-cm when $T \to 0$.[14-16] In contrast, a typical insulator has a $\rho > 10^{10}$ $\Omega$-cm even at the room temperature. On the other hand, the low-temperature phase of $La_{1-x}Sr_xMnO_3$ and $La_{1-x}Ca_xMnO_3$ with x smaller than about 0.50 exhibit ferromagnetic properties and have an increasing $\rho$ with T, so that these manganites are described as ferromagnetic metals, despite that $\rho$ of theses manganites have an order between $10^0$ and $10^{-3}$ $\Omega$-cm,[5,14] which is much larger than the typical $\rho$ of metals of an order of $10^{-6}$ $\Omega$-cm at room temperature.

The widely used phenomenological double exchange model of Zener[17,18] and Anderson[19] and its variations are based on the co-existence of $Mn^{3+}$ and $Mn^{2+}$ ions or $Mn^{3+}$ and $Mn^{4+}$ ions in so called electron- or hole-doped manganites, respectively, which have been the principle approaches to explain electronic, magnetic properties, and CMR of these manganites. However, the validity of the mixed-valence picture of Mn ions is questionable for the following reasons. First, the electronegativity of Mn and O are 1.55 and 2.86 [20], respectively, which indicates that the Mn-O bond is not 100% ionic, so that Mn ions cannot have whole-number effective charges. Second, it can be shown that the total electrostatic energy, which includes the self-energies of the positive Mn ions and surrounding compensating negative charges, of the mixed valence state is more repulsive than the uniform valence state. Third, $Mn^{3+}$ and $Mn^{4+}$ ions have differing on-site electrostatic potentials, which should show different chemical shifts of the Mn $3d$ $t_{2g}$ and $e_g$ features in the Mn $K$- and $L_{3,2}$-edge x-ray absorption spectroscopy (XAS) spectra. In other words, the XAS spectra should show two sets of Mn $4p$ or $3d$ features. However, only one set of Mn $4p$ or $3d$ features has been observed in XAS measurements.[21,22] One may argue that $Mn^{3+}$ and $Mn^{4+}$ ions exchange rapidly, so that the two sets of $t_{2g}$ and $e_g$ features are not resolved in XAS measurements. However, in this case Mn $3d$ XAS features should be significantly broadened with respect to those of undoped $LaMnO_3$, which has also been excluded by XAS measurements.[21,22] Observations of charge ordering in manganites have been regarded as support of the mixed valence picture. However, Loudon *et al.* questioned the interpretation of the $Mn^{3+}$ and $Mn^{4+}$ charge ordering in $La_{1-x}Ca_xMnO_3$ and argued that it should be interpreted by charge density wave instead.[23] Although the mixed valence picture of Mn ions in doped manganites has been regarded by many researchers as well-established, we believe a valid physical picture should be compatible with all experimental evidences.

On the other hand, the first-principles calculation methods have been argued to fail for strongly correlated electronic systems[24] such as manganites. This argument was based on an analogy with the hydrogen molecule. If the hydrogen molecular orbital is constructed from the linear combination of orbitals of the two hydrogen atoms, the two-electron wavefunction contains the Heitler-London and p+H⁻ (a proton and an H⁻ ion) parts. The p+H⁻ part gives rise to erroneous Coulomb energy if the intra-atomic Coulomb energy is strong, i.e. strongly correlated. However, the first-principles methods do not use two-electron wavefunctions and do not have the p+H⁻ part, so that the argument against the first-principles methods is invalid. The first-principles methods use Bloch wave functions as the basis set. In principle, the Bloch states form a complete set, so that any localized and singly occupied state can be expanded in spin-polarized Bloch states. A localized state will show as a flat energy band. The first-principles methods have the advantage that the relevant intra-Coulomb energy, U, is automatically included in the Kohn-Sham effective potential. And spin-polarization allows the occupation of a single electron in a localized state.

## II. CALCULATION METHOD AND CRYSTAL STRUCTURES

The spin-polarized first-principles calculation method used in this study is the modified pseudofunction (PSF) method,[25,26] which uses local-spin-density approximation (LSDA) of von Barth and Hedin[27] and the linear theory of Andersen[28]. This method is similar to the linearized muffin-tin orbital method (LMTO). The main difference between LMTO and PSF methods is the basis set. The PSF method uses both spherical Hankel and Nuemann tailing functions for lower- and higher-energy states, respectively, while the LMTO method uses only the spherical Hankel tailing functions with a zero kinetic-energy parameter.

In the current spin-polarized formalism of the PSF method, the sub-matrices for majority-spin ($\uparrow$-spin) and minority-spin ($\downarrow$-spin) states in the total Hamiltonian matrix are diagonalized separately, i.e. the off-diagonal sub-matrices, which accounts for the explicit couplings between $\uparrow$-spin and $\downarrow$-spin states are ignored. However, couplings between $\uparrow$-spin and $\downarrow$-spin states are implicitly included through the spin-polarized LSDA exchange-correlation potentials. Therefore, in this approach, self-consistent iterations that give the final energy state depend on the spin-polarized starting potential. Consequently, it does not automatically give the spin-polarized ground state. In this study, A-type (A-AF), G-type (G-AF), C-type (C-AF) antiferromagnetic states and the ferromagnetic (F) state are considered. (The spin orders of these magnetic types are shown in Fig. 1 of Ref. 29.) The relative stabilities of these magnetic states are determined by the comparison of their total energies.

In order to compare total energies of different types of magnetic orders on the same footing, the unit cells are chosen to have the same size for all types of spin orders for a given composition. The experimental lattice parameters of different crystal structures of $La_{1-x}Ca_xMnO_3$ and $La_{1-x}Sr_xMnO_3$ are used.[30-35] For x=1.00, i.e. $CaMnO_3$ and $SrMnO_3$, the cubic unit cell (2a,2b,2c) with 40 atoms is used, and for $La_{0.75}Sr_{0.25}MnO_3$, the rhombohedral $R\bar{3}c$ structure without distortion with 20 atoms per unit cell is used. The lattice constants of $La_{0.775}Sr_{0.225}MnO_3$ given in Ref. 34 are used for $La_{0.75}Sr_{0.25}MnO_3$. For $LaMnO_3$ and manganites with other compositions of Sr and Ca, the orthorhombic Pnma structure with distortions and 40 atoms per unit cell is used. Bloch sums of PSFs are expanded in 9261, 13357, and 9765 plane waves for x=1.00, $La_{0.75}Sr_{0.25}MnO_3$ and other manganites, respectively, and the charge density and potential are expanded by 68921, 99937, and 72529 plane waves for x=1.00, $La_{0.75}Sr_{0.25}MnO_3$ and other manganites, respectively. The special $k$ point scheme of Monkhorst and Pack[37] with q=4 in each direction of the reciprocal vector is used to approximate the integration over the first Brillouin zone.

## III. RESULTS AND DISCUSSION
### (a) Magnetic states

The total energies per Mn ion, i.e. the total energy per unit cell divided by the number of Mn ions in the unit cell, of the most favorable antiferromagnetic state relative to that of the ferromagnetic state are shown in Table I. The total energy results show that $LaMnO_3$, $La_{0.50}Ca_{0.50}MnO_3$, $La_{0.50}Sr_{0.50}MnO_3$, $La_{0.25}Ca_{0.75}MnO_3$, and $La_{0.25}Sr_{0.75}MnO_3$ are A-type antiferromagnetic, $La_{0.75}Ca_{0.25}MnO_3$ and $La_{0.75}Sr_{0.25}MnO_3$ are ferromagnetic, and $CaMnO_3$ and $SrMnO_3$ are G-type antiferromagnetic. The A-type antiferromagnetic state of $La_{0.25}Sr_{0.75}MnO_3$, which is only slightly more favorable than the C-type antiferromagnetic state by 0.13eV per 40-atom unit cell, disagrees with the C-type antiferromagnetic state shown in the phase diagram of $La_{1-x}Sr_xMnO_3$.[35] This discrepancy may be due to the particular crystal structure used in this study. Except $La_{0.25}Sr_{0.75}MnO_3$, the calculated magnetic states of these manganites agree with those shown in the phase diagrams of $La_{1-x}Sr_xMnO_3$ and $La_{1-x}Ca_xMnO_3$.[35, 36]

Table I The total energies per Mn ion of the most favorable antiferromagnetic state relative to that of the ferromagnetic state for (a) $La_{1-x}Ca_xMnO_3$ and (b) $La_{1-x}Sr_xMnO_3$.

(a)

| x=0.00 | x=0.25 | x=0.50 | x=0.75 | x=1.00 |
|---|---|---|---|---|
| **-0.051eV** | +0.062eV | **-0.036eV** | **-0.035eV** | **-0.199eV** |
| (A-AF) | (G-AF) | (A-AF) | (A-AF) | (G-AF) |

(b)

| x=0.00 | x=0.25 | x=0.50 | x=0.75 | x=1.00 |
|---|---|---|---|---|
| **-0.051eV** | +0.084eV | **-0.022eV** | **-0.045eV** | **-0.153eV** |
| (A-AF) | (A-AF) | (A-AF) | (A-AF) | (G-AF) |

### (b) Electronic properties

The calculated partial densities of states (PDOS) with energy levels broadened by a Gaussian function with a width parameter of 0.02eV for A-type antiferromagnetic $LaMnO_3$ are shown in Fig. 1. The Fermi level ($E_F$) is chosen to be the zero energy. $LaMnO_3$ has an energy gap of about 0.3eV formed between Mn $\uparrow$-spin $e_g$ and $\downarrow$-spin $t_{2g}$ subbands. Note that $LaMnO_3$ is specified as an insulator in the phase diagrams of $La_{1-x}Sr_xMnO_3$ and $La_{1-x}Ca_xMnO_3$.[35, 36] The energy gap of $LaMnO_3$ has been reported to be 1.1eV by optical absorption measurements.[13] However, the optical energy gap may be different from the actual energy gap because photo absorption/emission obeys dipole-transition selection rule, i.e. $\Delta l=1$, where $l$ is the orbital angular momentum quantum number. Thus, optical absorption/emission concerns transition from $s$ to $p$ or $p$ to $s/d$ states, so that the calculated d-band gap cannot be compared with the optical gap. Besides, LDA/LSDA is well known to underestimate the energy gap.

The spin-polarized PDOS's of ferromagnetic $La_{0.75}Ca_{0.25}MnO_3$ and $La_{0.75}Sr_{0.25}MnO_3$ given in Figs. 2(a) and 3(a), respectively, show that the leading edge of the $\uparrow$-spin $e_g$ band is broadened and extend above $E_F$ up to about 2.5eV. The O $2p$ states near $E_F$ are found to be delocalized and spread above $E_F$. Another change is the lowering of the $\downarrow$-spin $t_{2g}$ band such that its leading edge touches $E_F$. The sharp leading edge of the $\downarrow$-spin $t_{2g}$ band indicates that these states immediately above $E_F$ are localized with a large effective mass, so that they are not dominant contribution to the conductivity. The observed semimetallic behavior is dominantly due to delocalized $\uparrow$-spin $e_g$ and O $2p$ states.

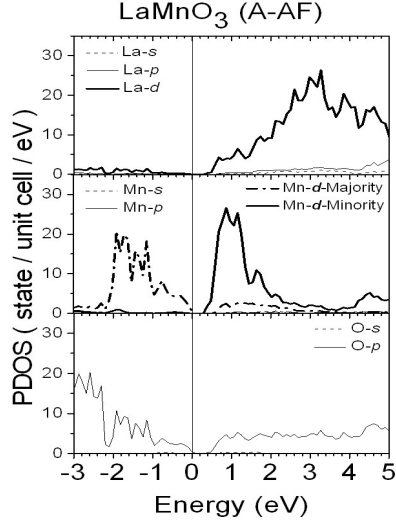

Fig. 1. The calculated partial densities of states (PDOS) for A-type antiferromagnetic $LaMnO_3$.

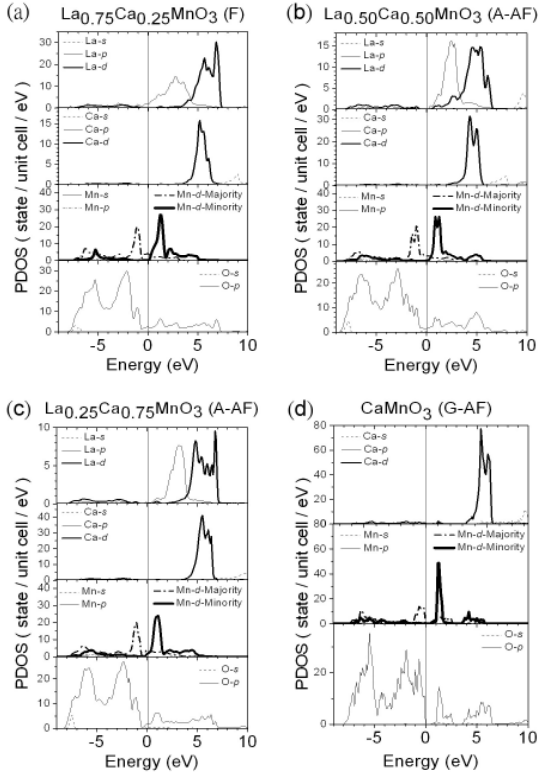

Fig. 2. Calculated partial densities of states (PDOS) for $La_{1-x}Ca_xMnO_3$ with x= (a) 0.25, (b) 0.50, (c) 0.75 and (d) 1.00.

The spin-polarized PDOS's of A-type antiferromagnetic $La_{0.50}Ca_{0.50}MnO_3$, $La_{0.25}Ca_{0.75}MnO_3$, $La_{0.50}Sr_{0.50}MnO_3$, and $La_{0.25}Sr_{0.75}MnO_3$ are shown in Figs. 2(b), 2(c), 3(b), and 3(c), respectively, and the spin-polarized PDOS's of G-type antiferromagnetic $CaMnO_3$ and $SrMnO_3$ are shown in Figs. 2(d) and 3(d), respectively. These figures show a trough, or pseudo-gap, located at $E_F$ which is formed by the overlapping of the leading edges of $\uparrow$-spin $e_g$ and

$\downarrow$-spin $t_{2g}$ Mn 3d subbands. The overall densities of delocalized $\uparrow$-spin $e_g$ and O 2p states in the vicinity of $E_F$ are reduced and the leading edge of the sharp $\downarrow$-spin $t_{2g}$ subband moves away from $E_F$ for large x. The phase diagrams describe $La_{1-x}Sr_xMnO_3$ and $La_{1-x}Ca_xMnO_3$ for x>~0.5 as insulators. Since $\rho$ has an order of only $10^2$ $\Omega$-cm at T$\to$0 for $La_{0.3}Ca_{0.7}MnO_3$,[16] the "insulator" behavior cannot be attributed to a real energy gap. The pseudo-gaps or troughs in the densities of states shown in Figs 2(b), 2(c), 3(b) and 3(c) are plausible explanation of the simultaneous existence of a relatively small $\rho$ at T$\to$0 and a decreasing $\rho$ with T, because the sharp rise of the density of $\downarrow$-spin $t_{2g}$ states and other states above $E_F$ corresponds to a sharp increase of carrier concentration with T, which increases the conductivity and decreases the resistivity. In fact, the decrease of $\rho$ with T in association with a pseudo-gap is a characteristic of quasicrystals.[38-41] The calculated electronic properties suggest that whether the manganite is metal-like or insulator-like depends on a detailed balancing between delocalized Mn $\uparrow e_g$ and O 2p states, which are metal-like, and the localized Mn $\downarrow t_{2g}$ states just above $E_F$, which are insulator-like.

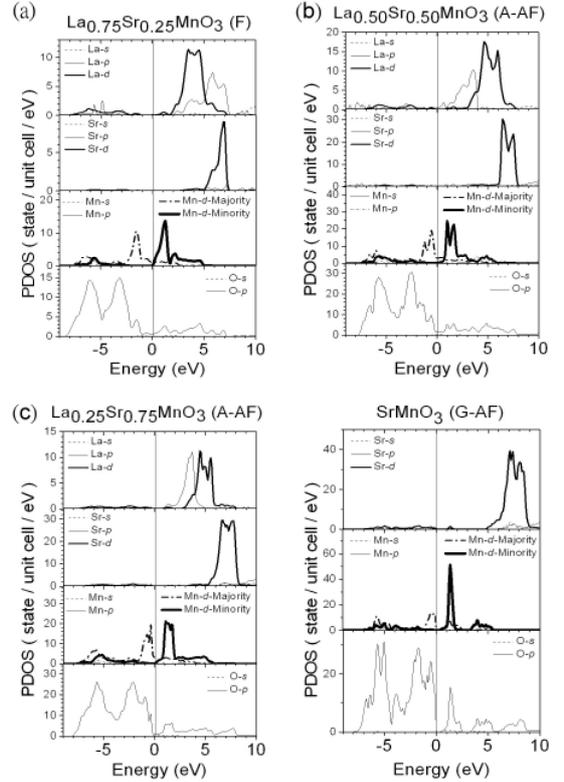

Fig. 3. Calculated partial densities of states (PDOS) for $La_{1-x}Sr_xMnO_3$ with x= (a) 0.25, (b) 0.50, (c) 0.75 and (d) 1.00.

**(c) Magnetic properties**

The calculated effective charges of Mn ions are ranged between ~+0.55e (for x=0) and ~+1.8e (for x=0.50), which are not +3e or +4e depicted in the double exchange model. The variation of the Mn effective charge is due to the subtle charge transfer in these quaternary compounds. The calculated average spin magnetic moments of Mn ions are 1.73, 1.66, 1.48, 1.41, and 1.18$\mu_B$ for $La_{1-x}Ca_xMnO_3$ with x=0.00, 0.25, 0.50, 0.75, and 1.00, respectively, and are 1.73,

1.74, 1.48, 1.40, and 1.23$\mu_B$ for La$_{1-x}$Sr$_x$MnO$_3$ with x=0.00, 0.25, 0.50, 0.75, and 1.00, respectively. These spin magnetic moments are significantly less than that of a free Mn atom of 2.5$\mu_B$ due to the broadening of the Mn 3$d$ bands as shown in Figs. 1-3.

The O- or group-VIB-anion mediated super-exchange coupling between adjacent Mn spins has been known to favor antiferromagnetism[42] and the mediation by itinerant or delocalized states in the vicinity of E$_F$ is known to favor ferromagnetism, for example the RKKY theory[43-45]. For pure LaMnO$_3$, the energy gap prevents ↑-spin $e_g$ states from mediating the ferromagnetic coupling, so that the O mediated super-exchange coupling dominates and the material is antiferromagnetic. In contrast, ↑-spin $e_g$ band extends above E$_F$ for sufficiently Sr or Ca doped manganites as shown in Figs. 2(a) and 3(a) for La$_{0.75}$Ca$_{0.25}$MnO$_3$ and La$_{0.75}$Sr$_{0.25}$MnO$_3$, respectively. The delocalized ↑-spin $e_g$ states near E$_F$ enhance delocalized-state mediated Mn-Mn spin couplings, so that the manganite becomes ferromagnetic. For G-AF CaMnO$_3$ and SrMnO$_3$ (x=1.0), there are energy gaps similar to LaMnO$_3$. For A-AF La$_{0.25}$Ca$_{0.75}$MnO$_3$ and La$_{0.25}$Sr$_{0.75}$MnO$_3$ (x=0.75), there is a trough near E$_F$ in the PDOS's of Mn ↑-spin $e_g$ subband, which indicates a reduction of the delocalized-state mediated ferromagnetic spin coupling. For A-AF La$_{0.50}$Ca$_{0.50}$MnO$_3$ and La$_{0.50}$Sr$_{0.50}$MnO$_3$ (x=0.50), the PDOS of Mn ↑-spin $e_g$ subband near E$_F$ is similar to that of the ferromagnetic x=0.25 case; the transition to the antiferromagnetic state is due to enhanced O mediated super-exchange coupling. According to Larson et al., the O mediated super-exchange integral is proportional to $\Delta E^{-3}$.[42] $\Delta E$ is the energy difference between the empty $d$ band and the valence band maximum, which is the highest occupied anion state that is not delocalized. $\Delta E$'s are 2.2eV, 1.5eV and 1.5eV, respectively, for x=0.25, 0.50 and 0.75 for La$_{1-x}$Sr$_x$MnO$_3$ and are 2.3eV, 1.8eV and 1.7eV, respectively, for x=0.25, 0.50 and 0.75 for La$_{1-x}$Ca$_x$MnO$_3$. $\Delta E$'s are smaller for x=0.50 and 0.75 than for x=0.25, which show that the O mediated super-exchange indeed is enhanced for x=0.50 and 0.75 relative to that for x=0.25. Thus, the present PDOS result explains why La$_{1-x}$Sr$_x$MnO$_3$ and La$_{1-x}$Ca$_x$MnO$_3$ become anti-ferromagnetic when x>~0.50. Based on calculated PDOS's, one can conclude that whether the manganite is ferromagnetic or antiferromagnetic depends on a detailed balancing between ↑-spin-$e_g$-states mediated ferromagnetic coupling and O mediated super-exchange anti-ferromagnetic coupling.

In doped manganites, the double exchange model has been commonly quoted to explain CMR by magnetic-field induced lining up of neighboring Mn spins, which enables Mn-$e_g$-state mediated double exchange that greatly reduces resistivity.[17-19] However, this interpretation ignored the fact that ferromagnetic materials are formed by magnetic domains. Even at a zero applied magnetic field neighboring Mn spins are already lined up within a given magnetic domain. An applied magnetic field triggers only the lining up of the magnetic polarizations of domains.

At a magnetic field of 6 tesla used in CMR measurements[44], the energy changes due to spin- and orbital-field couplings have an order of $10^{-4}$ eV, which is too small to cause significant change in carrier concentrations and average mobility. Thus, CMR is not due to changes in the electronic energies, but may be due to some form of magnetic-field induced instability like the lining up of magnetic domains. At a zero applied magnetic field, the orientations of the magnetic polarizations of the magnetic domains are disordered. Let us assume that magnetic domain A has a ↑-spin orientation, while a neighboring magnetic domain B has a ↓-spin orientation. Under an applied electric field, Mn ↑-spin $e_g$ electrons in domain A acquire momenta and move, say, toward domain B. When these mobile ↑-spin electrons enter into domain B, they see abundant empty localized Mn ↑-spin $t_{2g}$ states, which are minority-spin states in domain B. These ↑-spin $t_{2g}$ states in domain B become traps for the moving ↑-spin electrons, which reduces conductivity. When a magnetic field is applied and the magnetic moment or spin of domain B is triggered into lining up with that of domain A, the minority-spin, now ↓-spin, $t_{2g}$ states in domain B no longer trap the moving ↑-spin electrons unless there exits a spin-flip mechanism that couples these states. Then, the resistance to the moving ↑-spin electrons is suppressed. If this is the case, the resistivity or magnetoresistance as a function of the applied magnetic field should show saturation behavior. Indeed, the measurements of the resistivity as a function of the applied magnetic field for La$_{0.67}$Sr$_{0.33}$MnO$_3$ show that the resistivity saturates at a magnetic field of ~5000Oe as shown in Fig. 4, which supports the explanation of CMR given above.

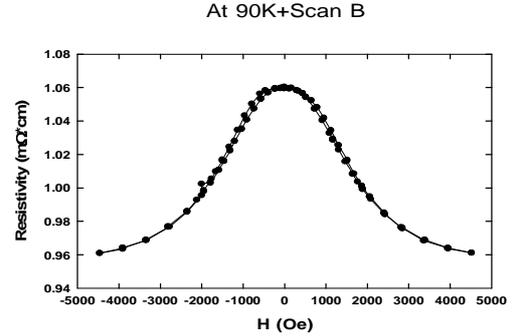

Fig. 4. The resistivity as a function of the applied magnetic field for La$_{0.67}$Sr$_{0.33}$MnO$_3$ at 90K.

## IV. CONCLUSION

Spin-polarized first-principles calculations show that the insulator behavior for La$_{1-x}$Sr$_x$MnO$_3$ and La$_{1-x}$Ca$_x$MnO$_3$ with x=0 and 1 is due to an energy gap, while the insulator-like behavior for 1>x>~0.5 is due to a pseudo-gap at E$_F$, which gives rise to a small $\rho$ at T→0 and a decreasing $\rho$ with T. The spin polarized PDOS's show that Sr and Ca induce delocalization of O 2$p$ and ↑-spin $e_g$ states, which renders these materials semimetallic and ferromagnetic. For x>~0.5, the decrease of the energy separation between empty Mn 3$d$ states and VBM enhances O mediated super-exchange coupling and in conjunction with the decrease of the number of delocalized states near E$_F$ renders these materials antiferromagnetic. The PDOS's also suggest that CMR is due to the conversion of carrier trapping Mn $t_{2g}$ states into non-trapping states by the applied-field-triggered lining up of domain polarization.

**Acknowledgement**

This work was supported by the National Science Council of Taiwan (contract number NSC 92-2112-M-110-013)